\documentclass[twocolumn,prd,showpacs,showkeys,preprintnumbers,amsmath,amssymb,floatfix]{revtex4}

\usepackage{graphicx}
\usepackage{dcolumn}
\usepackage{bm}

\begin{document}

\preprint{}

\title{New QCD Sum Rule for $D(0^+)$}

\author{Yuan-Ben Dai}
\email{dyb@itp.ac.cn} \affiliation{Institute of Theoretical Physics,
Chinese Academy of Sciences, P.O. Box 2735, Beijing 100080, China}

\author{Shi-Lin Zhu}
\email{zhusl@th.phy.pku.edu.cn} \affiliation{Department of Physics,
Peking University, Beijing 100871, China}

\begin{abstract}

We derive a new QCD sum rule for $D(0^+)$ which has only the $D\pi$
continuum with a resonance in the hadron side, using the assumption
similar to that has been successfully used in our previous work to
the mass of $D_s(0^+)(2317)$. For the value of the pole mass
$M_c=1.38 $ GeV as used in the $D_s(0^+)$ case we find that the mass
of $D(0^+)$ derived from this sum rule is significantly lower than
that derived from the sum rule with the pole approximation. Our
result is in agreement with the experimental dada from Belle.

\end{abstract}



\maketitle

\pagenumbering{arabic}

\section{ Introduction}
\label{sec1}

BaBar Collaboration discovered in 2003 a positive-parity scalar
charm strange meson $D_{sJ}(2317)$ with a very narrow width
\cite{babar}, which was confirmed by CLEO later \cite{cleo}. In the
same experiment CLEO \cite{cleo} observed the $1^+$ partner state at
$2460$ MeV. Since these two states lie below $DK$ and $D^\ast K$
threshold respectively, the potentially dominant s-wave decay modes
$D_{sJ}(2317) \to D_sK$ etc are kinematically forbidden. Thus the
radiative decays and isospin-violating strong decays become dominant
decay modes. Therefore they are very narrow.

The discovery of these two states has inspired great interest in
their nature in literature. The key point is to understand their low
masses. The $D_{sJ}(2317)$ mass is significantly lower than the
expected values in the range of $2.4-2.6$ GeV in quark models
\cite{qm}. The model using the heavy quark mass expansion of the
relativistic Bethe-Salpeter equation predicted a lower value $2.369$
GeV \cite{jin}, which is still 50 MeV higher than the experimental
data.

Later Belle \cite{belle} observed the wide $D(0^+)$ resonance with
mass $M_0=2.308\pm 0.0017\pm 0.0015\pm0.0028 $Gev and width around
$\Gamma=276$ MeV. Another puzzle arises from this result: why are
these two resonances nearly degenerate in mass while $D_s(0^-)$ is
100 MeV higher than $D(0^-)$?

The earlier results for the mass of $D_s(0^+)$ from QCD sum rules
are either significantly larger than the experimentally observed
mass of $D_{sJ}(2317)$ \cite{colangelo91} or consistent with it
within theoretical uncertainty but with significantly larger central
value \cite{dai}. The results from lattice QCD calculations are
similar, see \cite{bali} and \cite{dougall, soni}.

Recently, there have been two investigations on this problem using
sum rules in full QCD including the $O(\alpha_s)$ corrections. In
Ref. \cite{haya} the value of the charm quark pole mass $M_c=1.46
\text{GeV}$ is used and $0^+  \bar{c}s$ is found to be
$100-200\text{ MeV}$ higher than the experimental data. On the other
hand, in Ref. \cite{narison} $M_c\simeq 1.33 \text{GeV}$ is used and
the central value of the results for the $0^+$ $\bar{c}s$ mass is in
good agreement with the data. As commented by the author, the
uncertainty in the value of $M_c$ is large.

The difficulty with the $\bar c s$ interpretation leads many
authors to speculate that $D_{sJ}(2317)$ is a $\bar{c}qs \bar{q}$
four quark state \cite{cheng,barns} or a strong $D \pi $ atom
\cite{szc}. However, quark model calculations show that the mass
of the four quark state is much larger than the $0^+$ $\bar{c}s$
state \cite{vijande,zhang}. Furthermore, the four quark system has
two $0^+$ states. Only one has been found below 2.8 GeV in the
experimental search, consistent with the $\bar c s$
interpretation.

It was suggested in Ref. \cite{rupp} that the low mass of
$D_{sJ}(2317)$ could arise from the mixing between the $0^+$
$\bar{c}s$ state and the $DK$ continuum. It was pointed out in
\cite{dai} that in the formalism of QCD sum rules the physics of
mixing with $DK$ continuum resides in the contribution of $DK$
continuum in the sum rule and including this contribution should
render the mass of $D_s(0^+)$ lower.

Usually the contribution of the two-particle continuum is
neglected in the QCD sum rules. In a recent work \cite{dai06} we
argued that because of the large s wave coupling of $D_s(0^+)DK$
\cite{colangelo95,zhu} and the adjacency of the $D_s(0^+)$ mass to
the threshold this contribution may not be neglected. We
calculated this contribution under the assumption that the form
factor for the coupling of the scalar current to the two-particle
continuum in the low energy region is dominated by the bubble
diagrams formed by the coupling of the $0^+$ to the two particles.
We found that including this term in the sum rule indeed renders
the mass and decay constant of $D_s(0^+)$ significantly lower.
Using the pole mass value $M_c=1.38$ GeV the mass of $D_s(0^+)$ is
found to be in good agreement with experimental data.

In the present work we would like to show that the same assumption
for the form factor for the coupling of the scalar current to the
two-particle continuum also leads to the mass value of $D(0^+)$ in
agreement to the experimental data, thus, explains the nearly
degeneracy of these two states.

In Section \ref{sec2} we derive the $D\pi$ continuum contribution
and write down the full QCD sum rule for $D(0^+)$. The numerical
results of the mass of $D(0^+)$ are collected in Section \ref{sec3}.

\section{The QCD Sum Rule for the scalar charm meson}
\label{sec2}

We consider the scalar correlation function
\begin{equation}
\Pi (q^2) =i \int d^4x \exp[iqx]\langle 0| \textbf{T} \{J(x), J^\dag
(0) \}|0\rangle
\end{equation}
where $J(x)=\bar c(x) d(x)$ is the interpolating current for the
charged scalar charm meson. $\Pi(q^2)$ satisfies the following
dispersion relation
\begin{equation}
\Pi (q^2) ={1\over \pi} \int d s {\textbf{Im} \Pi (s) \over s-q^2 +
i\epsilon}\; .
\end{equation}
The leading terms of the imaginary part $\textbf{Im} \Pi (s)$ at the
quark gluon level and its $\alpha_s$ correction have been calculated
in \cite{reinders,narison}. After making the Borel transformation to
suppress the contribution of higher excited states and invoking the
quark-hadron duality, one arrives at the sum rule
\begin{eqnarray} \nonumber
\int dt {\textbf{Im}\Pi^{\text{H}} (t)\over \pi} \exp[-{t\over
M_B^2}] = {M_c^2\over (m_c-m_d)^2} \{
\int\limits_{(m_c+m_d)^2}^{\infty} dt   &\\ \nonumber \times
\exp[-{t\over M_B^2}] {1\over 8\pi^2} 3t(1-{M_c^2\over
t})^2  [ 1+{4\over 3}{\alpha_s\over \pi}G({M_c^2\over t})] &\\
\nonumber
 +M_c \langle \bar d d\rangle
\exp[-{M_c^2\over M_B^2}] + ({3\over 2}-{M_c^2\over
M_B^2}){\langle\alpha_s G^2\rangle \over 12\pi} &\\ +{M_c\over
2M_B^2}(1-{M_c^2\over 2M_B^2}) \langle g_s\bar d \sigma\cdot G
d\rangle \exp[-{M_c^2\over M_B^2}]
 \}
\end{eqnarray}
where $m_c$ and $m_d$ are the charm and down quark current mass.
The charm quark pole mass is defined as \cite{tara}
\begin{eqnarray}\nonumber
M_c = m_c(p^2) [ 1+\left({4\over 3}+ \ln{p^2\over M_c^2}
{\alpha_s\over \pi} \right)]\; .
\end{eqnarray}
$\langle \bar d d\rangle$ is the down quark condensate, $\langle
\alpha_s G^2\rangle$ is the gluon condensate, $\langle g_s\bar d
\sigma\cdot G d\rangle$ is the quark gluon mixed condensate. $M_B$
is the Borel mass. The radiative correction function reads
\begin{eqnarray} \nonumber
G(x)={9\over 4}+2\textbf{Li}_2(x)+ \log(x)\log(1-x) -{3\over
2}\log({1\over x}-1)&\\ \nonumber -\log(1-x) +x\log({1\over
x}-1)-{x\over 1-x} \log(x) \; .
\end{eqnarray}

The above equation is of the same form as that for $D_s(0^+)$. The
difference between the two cases is the following. $D_s(0^+)$ is a
very narrow resonance below the $DK$ threshold. Neglecting the
small iso-spin violating interaction it can be represented by a
pole term in the sum rule. On the other hand, $D(0^+)$ is a very
broad {\sl resonance} in the $D\pi$ spectra and there is no pole
term in the hadron spectra. In the traditional treatment of QCD
sum rules the very broad resonance is still approximated by a pole
term solely for simplicity and without justification. Moreover,
the remaining two-particle continuum is completely neglected.

According to this procedure the spectral density at the hadronic
level is taken to be the pole term plus the continuum starting
from some threshold which is identified with the QCD continuum.
\begin{eqnarray}\nonumber
{\textbf{Im}\Pi^{\text{H}} (t)\over \pi} = \left({f_{0}
M_{0}^2\over m_c-m_d}\right)^2 \delta (t-M^2_{0}) &\\
+\textbf{QCD continuum} \times \theta(t-s_0)
\end{eqnarray}
where $f_{0}$ is the vector current decay constant of $0^+$ $
\bar{c}d$ particle analogous to $f_\pi=132$ MeV, $M_{0}$ is the
mass of this particle, and $s_0$ is the continuum threshold above
which the hadronic spectral density is modelled by that at the
quark gluon level. The recent work \cite{narison} also uses the
above ansatz. The results from the above sum rule with $M_c=1.38$
GeV are shown in Fig. 1. This may not be a good approximation. The
importance of $D\pi$ continuum in the $D(0^+)$ channel was first
emphasized by Blok, Shifman and Uraltsev in Ref. \cite{shifman}
from the consideration of duality.
\begin{figure}[tbh]
\begin{center}
\scalebox{0.8} {\includegraphics{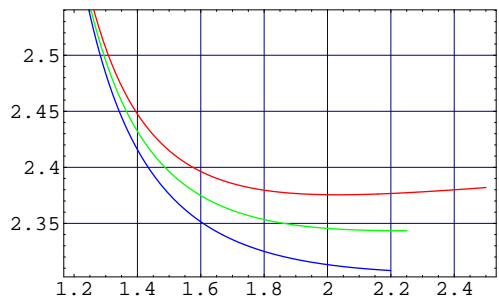}} \caption{$M_0$ vs
$M_B^2$ from the sum rule with pole approximation for $D(0^+)$.
Curves from top to bottom correspond to $s_0=8.0, 7.5, 7.0$
Gev$^2$ respectively.} \label{fig1}
\end{center}
\end{figure}

In our previous work \cite{dai06} it was pointed out that the
strong s wave coupling of the $D_s(0^+)$ with the two particle DK
state and the adjacency of the $D_s(0^+)$ mass to the DK continuum
threshold result in large coupling channel effect which
corresponds to the configuration mixing in the formalism of the
quark model. Therefore, the two-particle continuum can not be
neglected. In the present case of $D(0^+)$ the two-particle term
is the whole hadron spectra needed in the sum rule. We calculate
this term with the same assumption used in \cite{dai06}. Besides
giving a more accurate sum rule for the mass of $D(0^+)$, this
also constitutes a check for the approximation used in
\cite{dai06} for $D_s(0^+)$.

The contribution of the $D\pi$ continuum to the hadronic spectral
density reads
\begin{eqnarray}\nonumber
{\textbf{Im}\Pi^{\text{H}} (t)\over \pi} = & {3\over 32\pi^2}
\sqrt{1-{(M_D+m_\pi)^2\over t}} \sqrt{1-{(M_D-m_\pi)^2\over t}} \\
\nonumber
&\times |F(t)|^2 \theta(\sqrt{t}-M_D-m_\pi)\theta(s_0-t) \\
& +\textbf{QCD continuum}\times \theta(t-s_0) \; ,
\end{eqnarray}
where $F(t)$ is the form-factor defined by
\begin{eqnarray}
F(t)=\langle 0| \bar c (0) d(0) |D\pi\rangle
\end{eqnarray}
Similar to the approximation used in \cite{dai06}, from the large s
wave coupling of $D(0^+)D\pi$ we assume that in low energy region
$F(t)$ is dominated by the product of a factor proportional to the
propagator of the $D(0^+)$ and a factor from the final state
interaction represented by the chain of the bubble diagrams
generated from the $D(0^+)D\pi$ interaction as shown in Fig. 2.

\begin{figure}[tbh]
\begin{center}
\scalebox{0.5} {\includegraphics{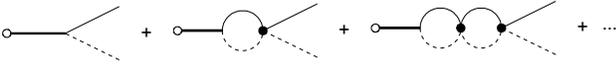}} \caption{Heavy,
light and dotted line represents $D(0^+)$, $D$ and $\pi$
respectively. Black circle represents the Born s wave amplitude of
$D\pi$ scattering. Blank circle represents the scalar current.}
\label{fig2}
\end{center}
\end{figure}

Let $g$ be the coupling constant in the effective $D(0^+)D\pi$
interaction lagrangian. The Born s wave amplitude in the figure
contains both a $t$ channel pole term $g^2\over t-M_0^2$ and a
crossing $s$ channel pole term
\begin{eqnarray}\nonumber
{g^2\over 2} \int\limits_{-1}^{+1} {{1\over s-M_0^2} d\cos \theta
}\; .
\end{eqnarray}
Using the relation between $s, t$ and the scattering angle $\theta$
one finds that the latter is an analytic function of $t$ with only a
short cut. Therefore, this term can be approximated by a pole form
$cg^2\over {t-t_0}$, where
\begin{eqnarray}\nonumber
t_0={{1\over 2} (2M_D^2+2m_\pi^2-M_{0}^2+{(M_D^2-m_\pi^2)^2\over
M_{0}^2})},
\end{eqnarray}
\begin{eqnarray}\nonumber
c=-{{2M_D^2+2m_\pi^2-M_{0}^2-{(M_D^2-m_\pi^2)^2\over
M_{0}^2}}\over{{(M_D^2-m_\pi^2)^2\over t_0} +t_0-2M_D^2-2m_\pi^2}}.
\end{eqnarray}
Different from the case of $D_s(0^+)$, the crossing term is actually
small for experimental values of the masses in the present case of
$D(0^+)$. But we shall still keep it in the formulae. In the chiral
lagrangian the interaction between $D$ and $\pi$ through exchanging
two pions is suppressed by $(E/{2\sqrt{2}\pi f_\pi})^4$ for low
energy $E$ of $\pi$. Hence its contribution can be neglected.
Summing the chain of the bubble diagrams, we have
\begin{eqnarray}\nonumber
F(t)={\lambda\over {(t-M_0^2)[1-({1\over t-M_0^2} +{c\over
{t-t_0}}) \Sigma(t)]}}
\end{eqnarray}
where
\begin{eqnarray}\nonumber
\Sigma (q^2) = -{3\over 2}ig^2\int {d^4k \over (2\pi)^4} {1\over
k^2-m_\pi}{1\over (q-k)^2-M_D^2}&\\ +\textbf{subtraction}
\end{eqnarray}
From Cutkosky's cutting rule,
\begin{eqnarray}
\textbf{Im}\Sigma(t)={3g^2\over 16\pi}{k(t)\over \sqrt{t}}\theta
(\sqrt{t}-M_D-m_\pi)\; ,
\end{eqnarray}
where
\begin{eqnarray}\nonumber
k(t)={\sqrt{\left(t-(M_D+m_\pi)^2\right)\left(t-(M_D-m_\pi)^2\right)}\over
2\sqrt{t}}
\end{eqnarray}
and we have taken into account two
intermediate states of different charge. From the dispersion
relation,
\begin{eqnarray} \nonumber
\Sigma(t)=\Sigma(M_0^2) +{1\over \pi}
\int\limits_{(M_D+m_\pi)^2}^{\infty} dt' {3g^2\over
16\pi}{k(t')\over \sqrt{t'}} {(t-M_0^2)\over (t'-M_0^2)(t'-t)}
\end{eqnarray}
where $M_0$ is the renormalized mass of the $D(0^+)$ meson. The mass
 renormalization condition leads to
$\textbf{Re}\Sigma(M_0^2)=0$. Therefore we have
\begin{eqnarray}\nonumber
& F(t)= \lambda \{(t-M_0^2-{3g^2\over 16\pi^2}(1+{c(t-M_0^2)\over {t-t_0}}) \\
\nonumber &\times  \textbf{P} \int\limits_{(M_D+m_\pi)^2}^{\infty}
dt'{k(t')\over \sqrt{t'}}  {(t-M_0^2)\over (t'-M_0^2)(t'-t)})\\
&-i{3g^2\over 16\pi}(1+{c(t-M_0^2)\over {t-t_0}}){k(t)\over
\sqrt{t}}\theta(\sqrt{t}-M_D-m_\pi)\}^{-1} \; ,
\end{eqnarray}
where $\textbf{P}$ denotes the principal part of integration.

The unknown constant $\lambda$ in $F(t)$ can be determined by
applying the soft-pion theorem to the extrapolated value of the
matrix element $\langle 0| \bar c (0) d(0) |D\pi\rangle$ at
$t=M_D^2$ with the result
\begin{eqnarray}\nonumber
F(M_D^2)={f_DM_D^2\over f_\pi (m_c+m_d)}\; .
\end{eqnarray}
Hence we have
\begin{eqnarray}\nonumber
\lambda =& {f_DM_D^2 \over f_\pi (m_c+m_d)}(M^2_D-M_0^2)
[1-{3g^2\over 16\pi^2}(1+{c(M^2_D-M_0^2)\over {M^2_D-t_0}}) \\
\nonumber & \times \textbf{P} \int\limits_{(M_D+m_\pi)^2}^{\infty}
dt'{k(t')\over \sqrt{t'}}  {1\over (t'-M_0^2)(t'-M_D^2)}]
\end{eqnarray}

\section{The Mass of $D(0^+)$}\label{sec3}

In the numerical analysis, we use $m_d=8$ MeV, $m_c=1.18$ GeV,
$M_c=1.38$ GeV, $\langle \bar d d\rangle =-(0.243)^3$ GeV$^3$,
$\langle \bar d g_ \sigma\cdot G d\rangle =0.8 \text{GeV}^2*\langle
\bar d d\rangle$, $\langle \alpha_s G^2\rangle =0.06$ GeV$^4$,
$\Lambda_{\text{QCD}}=0.325$ GeV, $M_D=1.869$ GeV, $m_\pi=0.140$
GeV, $f_D=0.2 $ GeV \cite{fd}, $f_\pi=0.132$ GeV \cite{pdg}.

The $g$ value has not been determined very well theoretically. It
was found to be in the interval $g=7.5-5.1 $ GeV in Ref.
\cite{colangelo95}. Inclusion of the contribution of $D\pi$
continuum in the sum rule analysis of the scalar current channel
may lower the $g$ value. Since the uncertainty is large, we
determine it from the experimental width $\Gamma$ of $D(0^+)$
through the equation
\begin{eqnarray}\nonumber
g^2=\Gamma[{3\sqrt{\left(M_0^2-(m_D+m_\pi)^2)\right)\left(
M_0^2-(m_D-m_\pi)^2\right)}\over {32\pi M_0^3}}]^{-1} \; ,
\end{eqnarray}
and use the experimental central value $\Gamma=276$ MeV
\cite{belle} as the input. For the experimental central value of
the $D(0^+)$ mass $M_0=2.308$ GeV, $g=8.09$ GeV. Notice that,
since we can not do wave function renormalization for $D(0^+)$ at
the resonance energy $M_0$, the coupling constant $g$ defined in
this work is not exactly the correspondence of that defined in
Ref. \cite{dai06} for $D_s(0^+)$.

\begin{figure}[tbh]
\begin{center}
\scalebox{0.8} {\includegraphics{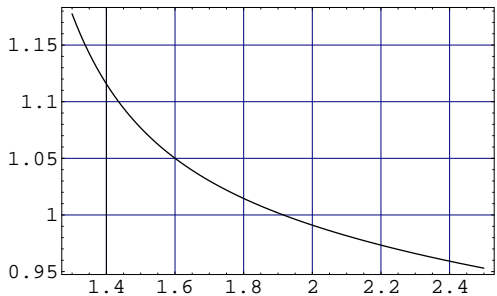}} \caption{The
variation of the ratio of the LHS to the RHS of the sum rule with
$M_B^2$ when $s_0 = 8.0 \text{GeV}^2$, $M_0=2.297$ GeV.}
\label{fig3}
\end{center}
\end{figure}

\begin{figure}[tbh]
\begin{center}
\scalebox{0.8} {\includegraphics{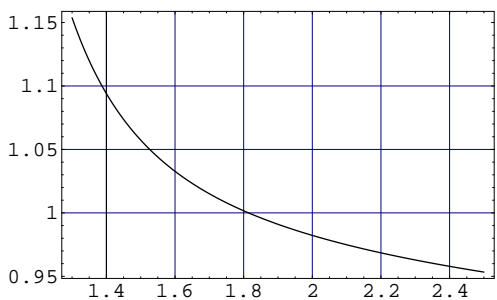}} \caption{The
variation of the ratio of the LHS to the RHS of the sum rule with
$M_B^2$ when $s_0 = 7.5 \text{GeV}^2$, $M_0=2.285$ GeV.}
\label{fig4}
\end{center}
\end{figure}

The good convergence of the OPE series and dominance of the $D\pi$
term over the QCD continuum beyond $s_0$ constrain the Borel mass
in a region depending on $M_c$ and $s_0$. For $M_c=1.38$ GeV,
$s_0=8.0,7.5,7.0$ GeV$^2$, $M_B^2\in [1.15, 2.5], [1.15, 2.25],
[1.15, 2.2]$ GeV$^2$ respectively. Since the hadron side of the
sum rule depends on the unknown value $M_0$ in a complicated way,
we have to input a "trial" value of $M_0$ and require that the
ratio of the hadron side (the left hand side or LHS) to the
quark-gluon side (the right hand side or RHS) of the sum rule is
equal to 1 in the middle of the working range of $M_B^2$. The
results are shown in Fig. 3, 4, 5 for $s_0=8.0, 7.5, 7.0$
respectively.

Comparing these results to those obtained in the pole
approximation shown in Fig. 2 one finds that the results for the
mass of $D(0^+)$ are lower by $80-40$ MeV. For $M_c=1.38$ GeV,
$s_0=8.0-7.0$ GeV$^2$ we find $M_0=2.30-2.27$ GeV in agreement
with the experimental data. The curve is very sensitive to the
trial value of $M_0$, hence the uncertainty of our results from
different values of $M_B$ for fixed values of other parameters are
small. Therefore, with the same approximation for the two-particle
continuum and same values of $M_c$ and $s_0$ we obtain
simultaneously the correct values of the masses of the narrow
resonance $D_s(0^+)$ below the threshold and the broad resonance
$D(0^+)$ above the threshold.

Under our approximation for the two-particle continuum and
acceptable value of $g$, there is no solution for the sum rule for
$D(0^+)$ containing a below-threshold pole term and the
two-particle continuum in the hadron side as that used in the
$D_s(0^+)$ case.

\begin{figure}[tbh]
\begin{center}
\scalebox{0.8} {\includegraphics{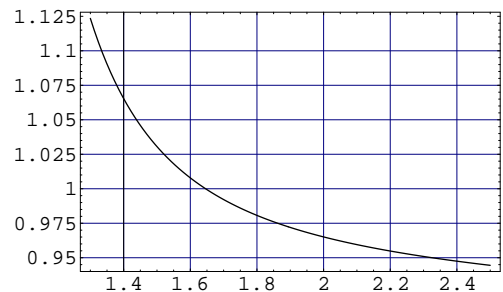}} \caption{The
variation of the ratio of the LHS to the RHS of the sum rule with
$M_B^2$ when $s_0 = 7.0 \text{GeV}^2$, $M_0=2.270$ GeV. }
\label{fig5}
\end{center}
\end{figure}

{\bf Acknowledgments:} S.L.Z. was supported by the National Natural
Science Foundation of China under Grants 10375003 and 10421503, and
Key Grant Project of Chinese Ministry of Education (NO 305001).

\end{document}